\documentstyle[12pt]{article}
\newcommand\be{\begin{equation}}
\newcommand\ee{\end{equation}}
\newcommand\bea{\begin{eqnarray}}
\newcommand\eea{\end{eqnarray}}

\newcommand{\fatalpha}{{\bf \alpha \kern -0.44em \alpha}}
\newcommand{\fatsigma}{{\bf \sigma \kern -0.54em \sigma}}
\newcommand{\tpchi}{{\bf \chi \kern -0.35em \chi}}
\newcommand{\llambda}{{\bf \lambda \kern -0.45em \lambda}}

\bibliography{plain}
\title{\bf Correlation dynamics of three spin under a classical dephasing environment }\vspace{20mm}

\author{ M. Mahdian$^a$\thanks{Mahdian@tabrizu.ac.ir},
          R. Yousefjani$^b$\thanks{R.yousefjany@uok.ac.ir}
          and S. Salimi$^b$\thanks{shsalimi@uok.ac.ir}\\
          {$^a$\footnotesize \emph{Department of Theoretical Physics and Astrophysics, University of Tabriz,}}\\
           {\footnotesize \emph{P.O.Box 51664 , Tabriz , Iran}.}\\
            {$^b$\footnotesize \emph{Department of Physics, University of Kurdistan,}}\\
          {\footnotesize \emph{P.O.Box 66177-15175 , Sanandaj,
           Iran}.}
        }

\pagebreak

\vspace{20mm}
\begin{document}
\maketitle \vspace{10mm}
\begin{abstract}
By starting from the stochastic Hamiltonian of the three correlated
spins and modeling their frequency fluctuations as caused by
dephasing noisy environments described by Ornstein-Uhlenbeck
processes, we study the dynamics of quantum correlations, including
entanglement and quantum discord. We prepared initially our open
system with Greenberger-Horne-Zeilinger or W state and present the
exact solutions for evolution dynamics of entanglement and quantum
discord between three spins under both Markovian and non-Markovian
regime of this classical noise. By comparison the dynamics of
entanglement with that of quantum discord we find that entanglement
can be more robust than quantum discord against this noise. It is
shown that by considering non-Markovian extensions the survival time
of correlations prolong.
\end{abstract}

\section{Introduction}
The unavoidable interaction of any realistic quantum system with its
environment can destroy coherence between the states of a quantum
system. This is decoherence, which can be introduced by various
noisy models. Since, this phenomenon is deemed as one of the main
obstacles to the realization of quantum information processing,
recently, much attention has been paid on influence of it in quantum
process \cite{7,8,9}. Decoherence dynamics of entanglement and
quantum discord, as two very useful resources to perform various
quantum tasks, under the influence of environmental noises has been
extensively discussed \cite{1,2,3,4,5,6,13,14,15,16}. When the
environment correlation time is much shorter than the relaxation
time of the system, i.e., environment has no memory, the Markovian
approximation can be used to quantified the decoherence dynamics
\cite{10,11,12}. This approximation simplifies greatly the
mathematics to solve the dynamics of open quantum system, but, in
truth every environment is non-Markovian, i.e., with memory. The
non-Markovian effect is a kind of dynamic backaction effect on the
system due to pronounced memory effect of environment which could
compensate the lost coherence of quantum system. Preservation of
entanglement and discord can be caused by this characteristic of
non-Markovian decoherence.
\\Many investigator have lately used the well-known classical Ornstein-Uhlenbeck (OU) process as a model of non-Markovian decoherence \cite{17,18,19}.
The OU process has a long history in physics and plays a central
role in the mathematical descriptions of Brownian motion and Johnson
noise \cite{20,21}. In this letter, we present the exact solutions
for evolution dynamics of entanglement and quantum discord between
three spins, initially described by pure Greenberger-Horne-Zeilinger
(GHZ) or W state, under the classical OU noise. As expected, the
life of entanglement and quantum discord prolong due to the feedback
dynamical of non-Markovian decoherence. The aim of this paper is to
compare the robustness of entanglement with that of quantum discord
against to the OU decoherence. We find that entanglement can be more
strong than quantum discord under this noise.
\\The structure of this article is as follows: First we describe the model of OU decoherence and two correlation measures which we used for this study. In next section we present our results for entanglement and quantum discord between three spins with initial GHZ and W state. Summary of this letter are given in the last section.


\section{The Model}
\subsection{The Ornstein-Uhlenbeck decoherence}
The random fluctuations of a spin-frequency which are sufficiently
small, can approximately be described as an OU process \cite{17,22}.
Based on this, the stochastic Hamiltonian of the three correlated
spins with random fluctuations frequency
\begin{eqnarray}
H_{tot}(t)=\sum_{i=1}^{3}\frac{1}{2}\Omega_{i}(t)\sigma_{z}^{(i)},
\end{eqnarray}
can be assumed as the Hamiltonian of interaction between system and
the local OU noise. Under this condition, $\Omega_{i}(t)$ which is
the independent transition frequency of the $i$-th spin, describe
the local OU noise with the statistical mean value properties as
\begin{eqnarray}
M[\Omega_{i}(t)]&=&0, \cr
M[\Omega_{i}(s)\Omega_{i}(t)]&=&\beta(s-t)=\frac{1}{2}\Gamma_{i}\gamma
e^{-\gamma |s-t|}\,.
\end{eqnarray}
In this classical noise, $\gamma$ is the bandwidth of the noise,
which is related to the correlation time as
$\tau_{c}=\frac{1}{\gamma}$. Figuratively, the correlation time can
be deemed as the time scale over which the environment has memory
and after it the environment is back to equilibrium and effectively
has no memory. The coupling strength between the $i$-th spin and the
environment is given by $\Gamma_{i}$, ($i=1,2,3$), which for
simplicity we assumed  $\Gamma_{i}=\Gamma$ for all spins.
\\The time evolution of the total system can be calculated explicitly from
\begin{eqnarray}
\rho_{tot}(t)= U(t)\rho_{tot}(0)U^{\dag}(t),
\end{eqnarray}
where the stochastic unitary operator $U(t)$ is the explicit
solution for the stochastic Schrodinger equation as
\begin{eqnarray}
U(t)=e^{-i\int_{0}^{t}H_{tot}(s)ds}\,.
\end{eqnarray}
Clearly, the unitary operator $U(t)$ is dependent on the noise. By
taking the ensemble average over the noise field, ( i.e., from the
statistical mean ) the reduced density matrix of open spins system
$\rho_{s}(t)$ is obtained
\begin{eqnarray}
\rho_{s}(t)=M[\rho_{tot}(t)].
\end{eqnarray}
So, the dynamics of the system density matrix can be described in
terms of Kraus representation as
\begin{eqnarray}
\rho_{s}(t)= \sum_{i}K_{i}(t)\rho_{s}(0)K_{i}^{\dag}(t).
\end{eqnarray}
By using the fact that $Tr[\rho_{s}(t)]=1$, a condition on the Kraus
operators can be obtained as $\sum_{i}K_{i}^{\dag}(t)K_{i}(t)=I$. In
our case, the Kraus operators describing the interaction with the
local OU noise are given by
\begin{eqnarray}
K_{1}=F_{1} \otimes F_{1} \otimes F_{1}, \,\,\,\,\,\,\, K_{2}=F_{1}
\otimes F_{1} \otimes F_{2}, \,\,\,\,\,\,\, K_{3}=F_{1} \otimes
F_{2} \otimes F_{1}, \,\,\,\,\,\,\, K_{4}=F_{2} \otimes F_{1}
\otimes F_{1},\cr
 K_{5}=F_{1} \otimes F_{2} \otimes
F_{2},\,\,\,\,\,\,\, K_{6}=F_{2} \otimes F_{1} \otimes
F_{2},\,\,\,\,\,\,\, K_{7}=F_{2} \otimes F_{2} \otimes
F_{1},\,\,\,\,\,\,\, K_{8}=F_{2} \otimes F_{2} \otimes F_{2},\cr
\end{eqnarray}
where
\begin{eqnarray}
F_{1}=\left(\matrix{\mu(t)&0\cr 0&1}\right), \,\,\,\,\,\,
F_{2}=\left(\matrix{\nu(t)&0\cr 0&0}\right).
\end{eqnarray}
The parameters appearing in $F_{1}$ and $F_{2}$ operators are
\begin{eqnarray}
\mu(t)&=&\exp[\int_{0}^{t}\beta(s-t)d^{2}s]=\exp[\frac{-\Gamma}{2}\{t+\frac{1}{\gamma}(e^{-\gamma
t}-1)\}],\cr \nu(t)&=&\sqrt{1-\mu^{2}(t)}.
\end{eqnarray}
For sufficiently large values of bandwidth,
$\gamma\rightarrow\infty$, i.e., the correlation time
$\tau_{c}\rightarrow 0$, we get $\beta(s-t)=\Gamma \delta(s-t)$, and
hence the Markovian dynamic of OU noise is recovered. For this
limit, the factor of decoherence is reduced to
$\mu(t)\rightarrow\exp[\frac{-\Gamma t}{2}]$, hence, the coherence
decay rate determine by the coupling strength between the spins and
environment $\Gamma$. If the dynamics of system and environment is
such that the correlation function in Eq. (2) can not be replaced by
a delta function, (i.e., $\gamma\rightarrow 0$ ), the dynamics is
non-Markovian and memory effects of the environment have important
roles. In this case, by using the approximation $e^{-\gamma
t}\simeq1-\gamma t +\frac{1}{2}(\gamma t)^{2}$, the factor of
decoherence, can be expanded as $\mu(t)\rightarrow\exp[\frac{-1}{4}
\gamma \Gamma t^{2}]$.


\subsection{Measuring entanglement}
Since entanglement is conceived as a resource to perform various
tasks of quantum information processing \cite{23,24,25,26},
knowledge about the amount of entanglement in a quantum state is so
important. Indeed, awareness from the value of entanglement, means
knowing how well a certain task can be accomplished. The
quantification problem of entanglement only for bipartite systems in
pure states \cite{27} and two-qubit system in mixed state \cite{28}
is essentially solved. In multi-partite systems, even the pure state
case, this problem is not exactly solved and just lower bounds for
the entanglement have been proposed \cite{29,30,31,32}. Here, in
order to determination the exact minimum of entanglement between
three spins, we use the lower bound of concurrence for three-qubit
state which is recently suggested by Li et al. \cite{33}
\begin{eqnarray}
\tau_{3}(\rho)=\frac{1}{\sqrt{3}}(\sum_{j=1}^{6}(C_{j}^{12|3})^{2}+(C_{j}^{13|2})^{2}+(C_{j}^{23|1})^{2})^{\frac{1}{2}},
\end{eqnarray}
where $C_{j}^{12|3}$ is terms of the bipartite concurrences for
qubits $12$ and $3$ which is given by
\begin{eqnarray}
C_{j}^{12|3}=\max
\{0,\lambda_{j}^{12|3}(1)-\lambda_{j}^{12|3}(2)-\lambda_{j}^{12|3}(3)-\lambda_{j}^{12|3}(4)\}.
\end{eqnarray}
In this notation, $\lambda_{j}^{12|3}(\kappa)$, $(\kappa=1..4)$, are
the square nonzero roots, in decreasing order, of the non-Hermitian
matrix $\rho \tilde{\rho}_{j}^{12|3}$. The matrix
$\tilde{\rho}_{j}^{12|3}$ are obtained from rotated the complex
conjugate of density operator, $\rho^{*}$, by the operator
$S_{j}^{12|3}$ as $\tilde{\rho}_{j}^{12|3}=S_{j}^{12|3}\, \rho^{*}
\, S_{j}^{12|3} $. The rotation operators $S_{j}^{12|3}$ are given
by tensor product of the six generators of the group SO(4),
($L_{j}^{12}$), and the single generator of the group SO(2),
($L_{0}^{3}$) that is $S_{j}^{12|3}=L_{j}^{12} \otimes L_{0}^{3}$.
Since the matrix $S_{j}^{12|3}$ has four rows and columns which are
identically zero, so the rank of non-Hermitian matrix $\rho
\tilde{\rho}_{j}^{12|3}$ can not be larger than 4, i.e.,
$\lambda_{j}^{12|3}(\kappa)=0$ for $\kappa\geq 5$. The bipartite
concurrences $C^{13|2}$ and $C^{23|1}$ are defined in a similar way
to $C^{12|3}$.


\subsection{Measuring quantum discord}
Another kind of quantum correlation which is in general different
from entanglement has been designated as quantum discord \cite{34}.
This measure of quantum correlation, which is arising from the
difference between two quantum extensions of the classical mutual
information, proved its abilities as a fundamental resource for
quantum information tasks \cite{35,36,37,38}. Recently, many efforts
to generalization of quantum discord to multi-partite systems have
been made by different authors \cite{39,40}. The global measure of
quantum discord which is obtained by a systematic extension of the
bipartite quantum discord is the result of such efforts \cite{41}.
By using this measure one can quantifies the quantum discord of an
arbitrary multi-partite state. Under a set of von-Neumann
measurements as
\begin{eqnarray}
\Pi_{1}^{(l)}=\left(\matrix{\cos^{2}(\frac{\theta_{l}}{2})&
\,\,\,\,\,\,\,\,\,e^{i\varphi_{l}}\cos(\frac{\theta_{l}}{2})\sin(\frac{\theta_{l}}{2})\cr
e^{-i\varphi_{l}}\cos(\frac{\theta_{l}}{2})\sin(\frac{\theta_{l}}{2})&
\,\,\,\,\,\,\,\,\,\sin^{2}(\frac{\theta_{l}}{2})}\right), \cr \cr
\cr \Pi_{2}^{(l)}=\left(\matrix{\sin^{2}(\frac{\theta_{l}}{2})&
-e^{-i\varphi_{l}}\cos(\frac{\theta_{l}}{2})\sin(\frac{\theta_{l}}{2})\cr
-e^{i\varphi_{l}}\cos(\frac{\theta_{l}}{2})\sin(\frac{\theta_{l}}{2})&\cos^{2}(\frac{\theta_{l}}{2})}\right),
\end{eqnarray}
which rotated the direction of the basis vector of $l$-th spin with
$\theta_{l}\in[0, \pi)$ and $\varphi_{l} \in [0, 2\pi)$, the global
quantum discord has the form
\begin{eqnarray}
\emph{D}(\rho)&=& \min _{\{\theta_{l},\varphi_{l}\}}[S(\rho \|
\Phi(\rho))- \sum_{l=1}^{3}S(\rho_{(l)}\| \Phi^{(l)}(\rho_{l}))]\cr
&=& \min
_{\{\theta_{l},\varphi_{l}\}}[S(\Phi(\rho))-S(\rho)-\sum_{l=1}^{3}(S(\Phi^{(l)}(\rho_{l}))-S(\rho_{l}))].
\end{eqnarray}
In this notation, $S(\rho)=-Tr[\rho \log_{2}\rho]$ is the
von-Neumann entropy of the density matrix and $\Phi^{(l)}(\rho_{l})=
\Pi_{1}^{(l)} \rho_{l} \,\, \Pi_{1}^{(l)}+ \Pi_{2}^{(l)} \rho_{l}
\,\, \Pi_{2}^{(l)}$ is the reduce density matrix after performing
the measurement on the $l$-th spin. The matrix $\Phi(\rho)$ obtain
by carrying out the 8 projective measurements
\begin{eqnarray}
\Pi_{1}&=&\Pi_{1}^{(1)}\otimes \Pi_{1}^{(2)} \otimes\Pi_{1}^{(3)},
\,\,\,\,\,\,\,\,\,\, \Pi_{2}=\Pi_{1}^{(1)}\otimes \Pi_{1}^{(2)}
\otimes\Pi_{2}^{(3)}, \,\,\,\,\,\,\,\,\,\,
\Pi_{3}=\Pi_{1}^{(1)}\otimes \Pi_{2}^{(2)} \otimes\Pi_{1}^{(3)}, \cr
\Pi_{4}&=&\Pi_{2}^{(1)}\otimes \Pi_{1}^{(2)} \otimes\Pi_{1}^{(3)},
\,\,\,\,\,\,\,\,\,\, \Pi_{5}=\Pi_{1}^{(1)}\otimes \Pi_{2}^{(2)}
\otimes\Pi_{2}^{(3)}, \,\,\,\,\,\,\,\,\,\,
\Pi_{6}=\Pi_{2}^{(1)}\otimes \Pi_{1}^{(2)} \otimes\Pi_{2}^{(3)}, \cr
\Pi_{7}&=&\Pi_{2}^{(1)}\otimes \Pi_{2}^{(2)} \otimes\Pi_{1}^{(3)},
\,\,\,\,\,\,\,\,\,\, \Pi_{8}=\Pi_{2}^{(1)}\otimes \Pi_{2}^{(2)}
\otimes\Pi_{2}^{(3)},
\end{eqnarray}
on the three correlated spins as:
$\Phi(\rho)=\sum_{m=1}^{8}\Pi_{m}\,\rho \,\, \Pi_{m}$. In order to
eliminate the dependence of quantum discord on the measurement
operators, we must find the measurement basis that minimizes
$\emph{D}(\rho)$.
\\In the following we will investigated the evolution of entanglement and quantum discord, as two different kinds of the quantum correlation, between three correlated spin which initially described by pure GHZ or W state under the OU Markovian and non-Markovian decoherences.


\section{The dynamics of three correlated spins under the OU noise}
In this section we present time evolution of entanglement and
discord between three spins which is coupled with OU classical
noise. We assume the open system is initially prepared in
inequivalent class of pure three qubit state with maximally quantum
correlation which is known as GHZ and W states.


\subsection{ Initial GHZ state}
First, we suppose that the initial state of the system is pure GHZ
state
\begin{eqnarray}
|GHZ\rangle=\frac{1}{\sqrt{2}}(|000\rangle+|111\rangle).
\end{eqnarray}
According to Eq. (6), the reduced density matrix of the system under
the OU decoherence can be expressed as
\begin{eqnarray}
\rho_{GHZ}(t)=\frac{1}{2}\{|000\rangle\langle
000|+|111\rangle\langle 111|\}+
\frac{\mu^{3}(t)}{2}\{|000\rangle\langle 111|+|111\rangle\langle
000|\}.\cr
\end{eqnarray}
The concurrence of the above density matrix can be easily computed
as
\begin{eqnarray}
\tau_{3}(\rho_{GHZ}(t))=\mu^{3}(t)\tau_{3}(\rho_{GHZ}(0)).
\end{eqnarray}
Now, we turn to calculating the quantum discord from Eq. (13). By
tracing out any two spins of the density matrix (16), one can obtain
the reduced density matrices of the subsystems as
$\rho_{1}=\rho_{2}=\rho_{3}=\frac{I}{2}$. So,
$\sum_{l=1}^{3}(S(\Phi^{(l)}(\rho_{l}))-S(\rho_{l}))=0$. The
von-Neumann entropy of the $\rho_{GHZ}(t)$ can be computed as
\begin{eqnarray}
\hspace{-10mm}S(\rho_{GHZ}(t))=1-\frac{1+\mu^{3}(t)}{2}\log_{2}(1+\mu^{3}(t))-\frac{1-\mu^{3}(t)}{2}\log_{2}(1-\mu^{3}(t)).
\end{eqnarray}
After some calculation to find the desired measurement basis which
minimizes the quantum discord, we deduced that the eigenbasis of the
Pauli matrix $\sigma_{z}$ are the best measurement basis. Under this
local measurement we obtain $S(\Phi(\rho_{GHZ}(t)))=1$ and
\begin{eqnarray}
\hspace{-10mm}D(\rho_{GHZ}(t))=\frac{1+\mu^{3}(t)}{2}\log_{2}(1+\mu^{3}(t))+\frac{1-\mu^{3}(t)}{2}\log_{2}(1-\mu^{3}(t)).
\end{eqnarray}
In order to comparison the robustness of the entanglement (17) and
the quantum discord (19) against to OU decoherence, we plot these
quantities as a function of the dimensionless scale $\Gamma t$ for
both Markovian (a) and non-Markovian (b) regime in Fig. 1. It is
explicit that, entanglement can be survived under the Markovian and
non-Markovian conditions more than the quantum discord. Moreover,
observe that due to the influence of memory effect of environment
the entanglement and the quantum discord have longer life. In the
other words, one could imagine that non-Markovian effect is a key
factor to preserve quantum correlation since non-Markovian effect is
a kind of backaction effect which could compensate the lost
coherence of quantum system.


\subsection{ Initial W state}
When the three correlated spins is initially prepared in the W state
\begin{eqnarray}
|W\rangle&=&\frac{1}{2}(|100\rangle+|010\rangle+\sqrt{2}|001\rangle),
\end{eqnarray}
the density matrix dynamics, according Eq. (6), can be express as
\begin{eqnarray}
\rho_{W}(t)=\frac{1}{4}\left(\matrix{0&0&0&0&0&0&0&0\cr
0&2&\sqrt{2}\mu^{2}(t)&0&\sqrt{2}\mu^{2}(t)&0&0&0 \cr
0&\sqrt{2}\mu^{2}(t)&1&0&\mu^{2}(t)&0&0&0 \cr 0&0&0&0&0&0&0&0 \cr
0&\sqrt{2}\mu^{2}(t)&\mu^{2}(t)&0&1&0&0&0 \cr 0&0&0&0&0&0&0&0 \cr
0&0&0&0&0&0&0&0 \cr 0&0&0&0&0&0&0&0}\right).
\end{eqnarray}
From the lower bound of concurrence for three-qubit state Eq. (10)
for the above matrix we obtain
\begin{eqnarray}
\tau_{3}(\rho_{W}(t))=\mu^{2}(t)\tau_{3}(\rho_{W}(0)).
\end{eqnarray}
In order to determine quantum discord according to Eq. (13), we
first evaluate the one-spin density matrices representing the
individual subsystems by tracing out two spins. These reduce density
matrices have the flowing form
$\rho_{1}=\rho_{2}=\rho_{3}=\frac{1}{4}\{3|0\rangle\langle0|+|1\rangle\langle1|\}$.
Next, by using the same procedure done in previous section to find
the desired measurement basis which minimizes the quantum discord,
we obtain the minimum of quantum discord with adopt local
measurements in the $\sigma_{z}$ eigenbasis for any spin. Under such
measurements the state of the single spin is not induced therefor we
have $\sum_{l=1}^{3}(S(\Phi^{(l)}(\rho_{l}))-S(\rho_{l}))=0$ and
$S(\Phi(\rho_{W}(t)))=\frac{3}{2}$. Finally, by inserting the
von-Neumann entropy of the time-dependent reduced density matrix for
three correlated spins in Eq. (13), we obtain the flowing form for
quantum discord
\begin{eqnarray}
D(\rho_{W}(t))&=&\frac{-1}{4}(5+\mu^{2}(t))+\frac{1}{4}(1-\mu^{2}(t))\log_{2}(1-\mu^{2}(t))\cr
&+&\frac{1}{8}\{(3+\mu^{2}(t)-\sqrt{1-2\mu^{2}(t)+17\mu^{4}(t)})\cr
&&\log_{2}(3+\mu^{2}(t)-\sqrt{1-2\mu^{2}(t)+17\mu^{4}(t)})\cr
&+&(3+\mu^{2}(t)+\sqrt{1-2\mu^{2}(t)+17\mu^{4}(t)})\cr
&&\log_{2}(3+\mu^{2}(t)+\sqrt{1-2\mu^{2}(t)+17\mu^{4}(t)})\}.
\end{eqnarray}
Notice that we shall re-normalize the Eq. (23) in such a way that we
have $D(\rho_{W}(0))=1$. From the expressions of entanglement (22)
and quantum discord (23), it can be found that for
$\mu(t)\rightarrow0$ quantum correlation between three spins
disappear. In Fig. 2, the dynamics of these quantities for both
Markovian (a) and non-Markovian (b) dynamics are illustrated.
Clearly, robustness of entanglement versus OU Markovian and
non-Markovian decoherence is more than quantum discord.
\\Notice that, since the factor of decoherence $\mu(t)$ appears as quadratic in quantum correlation of W state, this state has life longer than GHZ state with cubic decoherence. Comparison of plots in Fig. 1 and Fig. 2 justifies this fact.


\section{Summary}
In summary, we have investigated the exact decoherence dynamics of
quantum correlations, including entanglement and quantum discord,
between three correlated spin with initial GHZ and W state under a
local dissipative OU process. By studying both Markovian and
non-Markovian regime of this classical noise, we have found that
under Markovian dynamics quantum correlation suffers sudden death,
while due to the influence of memory effect of environment in
non-Markovian dynamics, it has longer life. Moreover, comparison of
the survival time of entanglement and quantum discord make it clear
that against dephasing OU decoherence, entanglement is more robust
than quantum discord, because it decays exponentially while the
discord of the same state decays logarithmic. Also, our results have
shown that W sate has strong dynamics under decoherence than GHZ
state.


\begin{figure}
\includegraphics{Fig.1(a).eps}\vspace{7.9cm}
\end{figure}
\begin{figure}
\includegraphics{Fig.1(b).eps}\vspace{7.9cm}\caption{(Color online) Entanglement
(solid blue) and quantum discord (dashed red) dynamics of three
correlated spin with initial GHZ stat under OU decoherence. Here, we
have chosen the parameters $\Gamma=1$. Fig. (a) corresponds to
Markovian regime with $\mu(t)\rightarrow\exp[\frac{-\Gamma t}{2}]$
and Fig. (b) to non-Markovian regime with
$\mu(t)\rightarrow\exp[\frac{-1}{4} \gamma \Gamma t^{2}]$ and
$\gamma=0.01$.}
\end{figure}

\begin{figure}
\includegraphics{Fig.2(a).eps}\vspace{7.9cm}
\end{figure}
\begin{figure}
\includegraphics{Fig.2(b).eps}\vspace{7.9cm}\caption{(Color online) Entanglement
(solid blue) and quantum discord (dashed red) dynamics of three
correlated spin with initial W stat under OU decoherence. Here, we
have chosen the parameters $\Gamma=1$. Fig. (a) corresponds to
Markovian regime with $\mu(t)\rightarrow\exp[\frac{-\Gamma t}{2}]$
and Fig. (b) to non-Markovian regime with
$\mu(t)\rightarrow\exp[\frac{-1}{4} \gamma \Gamma t^{2}]$ and
$\gamma=0.01$.}
\end{figure}

\end{document}